\documentclass[a4paper, 10pt]{article}

\usepackage{ifxetex}
\ifxetex

\usepackage{polyglossia}
\setmainlanguage[variant = british]{english}
\usepackage{csquotes}

\else
\usepackage[utf8]{inputenc}
\fi

\usepackage[]{mciteplus}
\mciteSetBstSublistMode{n}
\mciteSetMidEndSepPunct{\quad$\bullet$\quad}{.}{\relax}

\usepackage{amsmath}

\newcommand{\ek}[1]{\begin{equation}#1\end{equation}}
\newcommand{\mek}[2]{\ek{\label{#2}#1}}
\newcommand{\mmek}[2]{\mek{\begin{split}#1\end{split}}{#2}}
\newcommand{\de}{\mathrm d}
\newcommand{\pe}{\partial}
\newcommand{\er}[1]{(\ref{#1})}
\newcommand{\tp}{{2\pi}}
\newcommand{\reals}{{\mathbf R}}
\newcommand{\mr}{\:\!}

\author{Jens Hoppe\footnote{KTH Royal Institute of Technology.}}
\title{Quasi-Static BMN Solutions}
\date{}

\begin{document}
\maketitle

\begin{abstract}
\noindent Classical solutions of membrane equations that were recently identified  as limits of matrix-solutions are looked upon from another angle. \\[6pt]
\end{abstract}
As extended objects can be described in many different ways, it is not always trivial to note when (and when not) differently looking expressions are describing the same (cp. e.\,g. \cite{arnlindEA-2013-1} where a rather non-trivial graph over the Clifford-torus is derived as a minimal surface in $S^3$, but is then shown to be, by an explicit, rather intricate reparametrisation, \emph{the} Clifford-torus; cp. also \cite{hoppe-1995, hoppe-2008} for surfaces involving time-dependent reparametrisations in rotations).

Recently, an interesting paper \cite{berensteinEA-2015} appeared in which exact solutions to finite $N$ matrix equations were constructed whose%%%% in rotations?
{} $N\to\infty$ limit will here be derived from a ``purely-continuum'' perspective: one way to solve the canonical equations of motion,

\mek{\ddot x_j = \{\{x_j, x_k\}, x_k\} - x_j + \tfrac32\epsilon_{jkl}\{x_k,x_l\},}{i}
corresponding to
\mmek{H = \tfrac12\int\left({\vec p\mr}^2 + \sum_{j=1}^3(x_j - \tfrac12\epsilon_{jkl}\{x_k,x_l\})^2\right)\de^2\varphi\\
=\tfrac12\int\left({\vec p\mr}^2 + {\vec x}^2 + \sum_{i<j}\{x_i, x_j\}^2 - \epsilon_{jkl}x_j\{x_k, x_l\}\right)\de^2\varphi,}{ii}
and also satisfying the constraint
\mek{\sum_{i=1}^3\{x_i,\dot x_i\} = 0}{iii}
is to assume $x_3(\tau, \varphi^1, \varphi^2) =: z(\varphi^1)$ to only depend on $\varphi^1 =: \varphi$, and to let
\mmek{ x_1 &= u(\varphi)\cos\Omega - v(\varphi)\sin\Omega\\
x_2 &= u(\varphi)\sin\Omega + v(\varphi)\cos\Omega\\
\Omega &= \Omega(\tau, \varphi, \varphi^2) := \varphi^2 + \omega(\varphi)\tau,}{iv}
i.\,e.
\mek{\begin{pmatrix}x_1\\ x_2\\ x_3\end{pmatrix} = \begin{pmatrix}\cos\Omega &-\sin\Omega &0\\\sin\Omega &\cos\Omega &0\\ 0 &0 &1\end{pmatrix}\begin{pmatrix}u\\ v\\ z\end{pmatrix}}{v}
to be quasi-static (the axially-symmetric shape of the surface described by \er{v} does not depend on $\tau$, the light-cone time).

Inserting \er{v} into \er{iii} one finds the requirement
\mek{\omega(\varphi)(u^2 + v^2)(\varphi) = \textrm{const} =: J;}{vi}
writing e.\,g. $x_a = R_{ab}u_{b}$ $(a,b = 1,2)$ one has
\mek{\sum_a\{x_a,\dot x_a\} = R_{ab}u_b'\omega R_{ac}''u_c - R_{ab}'u_bR_{ac}'(\omega'u_c + \omega u_c');}{vii}
then differentiate $R_{ab}R_{ac} = \delta_{bc}$ twice with respect to $\Omega$, and use $R_{\cdot\cdot}'' = - R_{\cdot\cdot}$ to arrive at \er{vi} (which is the analogue of eq. (36) of \cite{berensteinEA-2015}).

Inserting \er{v} into \er{i} and \er{ii} on the other hand gives (with $s := u^2 + v^2$, and using \er{vi})
\mek{\tfrac12s'' + \frac{J^2}{s^2} = 1 + {z'}^2 + 3z'}{viii}
\mek{(s(z' + \tfrac32))' = z}{ix}
\mek{\frac{E\{\vec u, z\}}\tp = \tfrac12\int\left(
\frac{J^2}s + (z - \tfrac12s')^2 + (1 + z')^2s
%(1 + {z'}^2){\vec u}^2 + 3z'{\vec u}^2 + z^2 + (uu' + vv')^2
\right)\de\varphi;}{x}
the easiest way to derive this is to note that $\dot x_1 = -\omega x_2$, $\dot x_2 = +\omega x_1$, $\pe_2x_x = x_1, \pe_2x_1 = -x_2$ and
\mmek{\{x_2,x_3\} &= -z'x_1 \ (= -x_3'x_1)\\
\{x_3, x_1\} &= -z'x_2 \ (= -x_3'x_2)\\
\{x_1, x_2\} &= x_1x_1' + x_2x_2' \ ( =\tfrac12s')}{xi}
-- which is (cp. \cite{arnlindEA-2007-1}) interesting in its own right.

Note that in \cite{berensteinEA-2015} (8)--(10) are derived as continuum limits of finite matrix expressions, and that (as for the finite $N$ case) the equations of motion of the reduced Hamiltonian coincide with what one gets when inserting \er{v} into the full equations \er{i}.

The shape of the curve whose rotation, cp. \er{v}, gives the surface in $\reals^3$, is given by $s$ (the square of the distance from the $z$-axis) and $z$, obtained by solving \er{viii} $+$ \er{ix}; e.\,g.
\mek{s(\varphi) = \frac1{z' + \frac32}\left(c + \int_0^\varphi z(\hat\varphi)\de\hat\varphi\right).}{xii}

For spherical surfaces it would actually suffice to calculate $s$ as a function of $z$. Choosing $z$ (instead of $\varphi$) as the variable is motivated also by \er{xi}, which after the transformation $\varphi \to \tilde\varphi := z(\varphi)$ reads
\mek{ \{x_2, x_3\} = -x_1,\ \{x_3, x_1\} = -x_2,\ \{x_1, x_2\} = \tfrac12 \pe_{\tilde\varphi}s;}{xiii}
the corresponding matrix equations (cp. \cite{arnlindEA-2010, *arnlindEA-2012, arnlindEA-2013-2}) would make $z = x_3$ canonical, resp.\ with integer-spaced diagonal entries.

Let me now comment on the ``BPS'' solutions
\mek{z' = -1 + \frac Js, \quad s' = 2z}{xiv}
of \er{viii} $+$ \er{ix} (cp. eq. (78)/(79) of \cite{berensteinEA-2015}, and the interpretation given there), which together with their finite $N$ counterparts were thoroughly discussed, and solved, already 10 years ago in \cite{bakEA-2005}.

Noting that \er{ii} may also be written as
\mmek{H &= \tfrac12\int p_3^2 + \tfrac12\int(p_1 \pm (x_2 - \{x_3,x_1\}))^2 + \tfrac12\int(p_2 \mp (x_1 - \{x_2, x_3\}))^2 \\
&+ \tfrac12\int(x_3 - \{x_1, x_2\})^2 \pm \int(x_1p_2 - x_2p_1) \pm \int x_3(\{p_1,x_1\} + \{p_2, x_2\}),}{xv}
with the last term being zero (because of \er{iii}), and the one before equalling $\pm 2\pi J$, one can identify the BPS equations for fixed $J$ to be
\mmek{\dot x_1 + x_2 - \{x_3, x_1\} = 0\\
\dot x_2 - x_1 + \{x_2, x_3\} = 0\\
\{x_1, x_2\} = x_3,\quad \dot x_3 = 0}{xvi}
which for $\dot x_1 = -\omega x_2$, $\dot x_2 = \omega x_1$ (cp. \er{iv}) and $\omega = J/s$ (cp. \er{vi}) become
\mmek{\{x_3, x_1\} = (1 - J/s)x_2\\
\{x_2, x_3\} = (1 - J/s)x_1}{xvii}
\mek{\{x_1, x_2\} = x_3,}{xviii}
implying in particular
\mek{\{\{x_3, x_1\}, x_1\} + \{\{x_3, x_2\}, x_2\} = -2x_3}{xix}
(cp.\ \cite{bakEA-2005, bak-2003, mikhailov-2002, hoppeEA-2007}). Using the same technique by which \er{xviii} $+$ \er{xix} were solved in \cite{bakEA-2005}, namely changing independent variables from $\varphi^1, \varphi^2$ to $x^1, x^2$ (cp. \cite{bordemannEA-1993}), giving
\mmek{\{f, x_1\} &= -\hat z\frac{\pe\hat f}{\pe x^2}\\
\{f, x_2\} &= \hat z\frac{\pe\hat f}{\pe x^1};}{xx}
one gets
\mmek{-\hat z\pe_2\hat z = (1 - J/s)x_2\\
-\hat z\pe_1\hat z = (1 - J/s)x_1}{xxi}
hence (cp. (82) of \cite{berensteinEA-2015})
\mek{x_1^2 + x_2^2 + x_3^2 - J\ln(x_1^2 + x_2^2) = \textrm{const.}}{xxii}
describing the shape of the membrane-solution.

%\printbibliography
\bibliographystyle{Style}
\bibliography{quasistatic}
\end{document}